\documentclass[aps,prl,twocolumn,showpacs,superscriptaddress]{revtex4}
\usepackage[latin1]{inputenc}
\usepackage[english]{babel}
\usepackage[T1]{fontenc}
\usepackage{bm,graphicx,graphics,amsmath,amssymb,epsfig,color}

\def \be {\begin{equation}}
\def \ee {\end{equation}}
\def \beA {\begin{eqnarray}}
\def \eeA {\end{eqnarray}}

\def \der {\partial}

\def \average#1{\left\langle #1 \right\rangle}

\begin{document}

\title{Designing a spin-Seebeck diode}

\author{Simone Borlenghi} 
\affiliation{Department of Materials and Nanophysics,  School of Information and Communication Technology, \\Electrum 229, Royal Institute of Technology, SE-16440 Kista, Sweden.}
\author{Weiwei Wang}
\affiliation{Engineering and the Environment, University of Southampton, SO17 1BJ Southampton, United Kingdom.}
\author{Hans Fangohr}
\affiliation{Engineering and the Environment, University of Southampton, SO17 1BJ Southampton, United Kingdom.}
\author{Lars Bergqvist} 
\affiliation{Department of Materials and Nanophysics,  School of Information and Communication Technology, \\Electrum 229, Royal Institute of Technology, SE-16440 Kista, Sweden.}
\affiliation{SeRC (Swedish e-Science Research Center), KTH, SE-10044 Stockholm, Sweden.}
\author{Anna Delin}
\affiliation{Department of Materials and Nanophysics,  School of Information and Communication Technology, \\Electrum 229, Royal Institute of Technology, SE-16440 Kista, Sweden.}
\affiliation{SeRC (Swedish e-Science Research Center), KTH, SE-10044 Stockholm, Sweden.}
\affiliation{Department of Physics and Astronomy, Uppsala University, Box 516, SE-75120 Uppsala, Sweden.}

\begin{abstract}
Using micromagnetic simulations, we have investigated spin dynamics in a spin-valve bi-layer in the presence of a thermal gradient. The direction and the intensity of the gradient allow
to excite the spin wave modes of each layer selectively. This permits to synchronize the magnetization precession of the two layers and to rectify the flows of energy and magnetization through the system.
Our study yields promising opportunities for applications in spin-caloritronics and nano-phononics devices.
\end{abstract}

\maketitle

The recent discovery of the Spin-Seebeck effect \cite{uchida08,uchida10} is at the core of spin-caloritronics \cite{bauer11}, an emerging field where the generation and control of spin currents
by a thermal gradient in nano-electronics and magnonic devices is in focus. In recent years, this field has been the object of intense investigation, 
yielding promising opportunity in energy efficient spintronics devices \cite{bauer11}. An essential step in this direction is the realisation of a diode that rectifies spin-current.
In the present paper, we investigate through micromagnetic simulations a realistic device, that behaves as a \emph{thermo-magnonic diode}, allowing the propagation of energy and magnetization
currents in one direction only.

\begin{figure}
\begin{center}
\includegraphics[width=7.0cm]{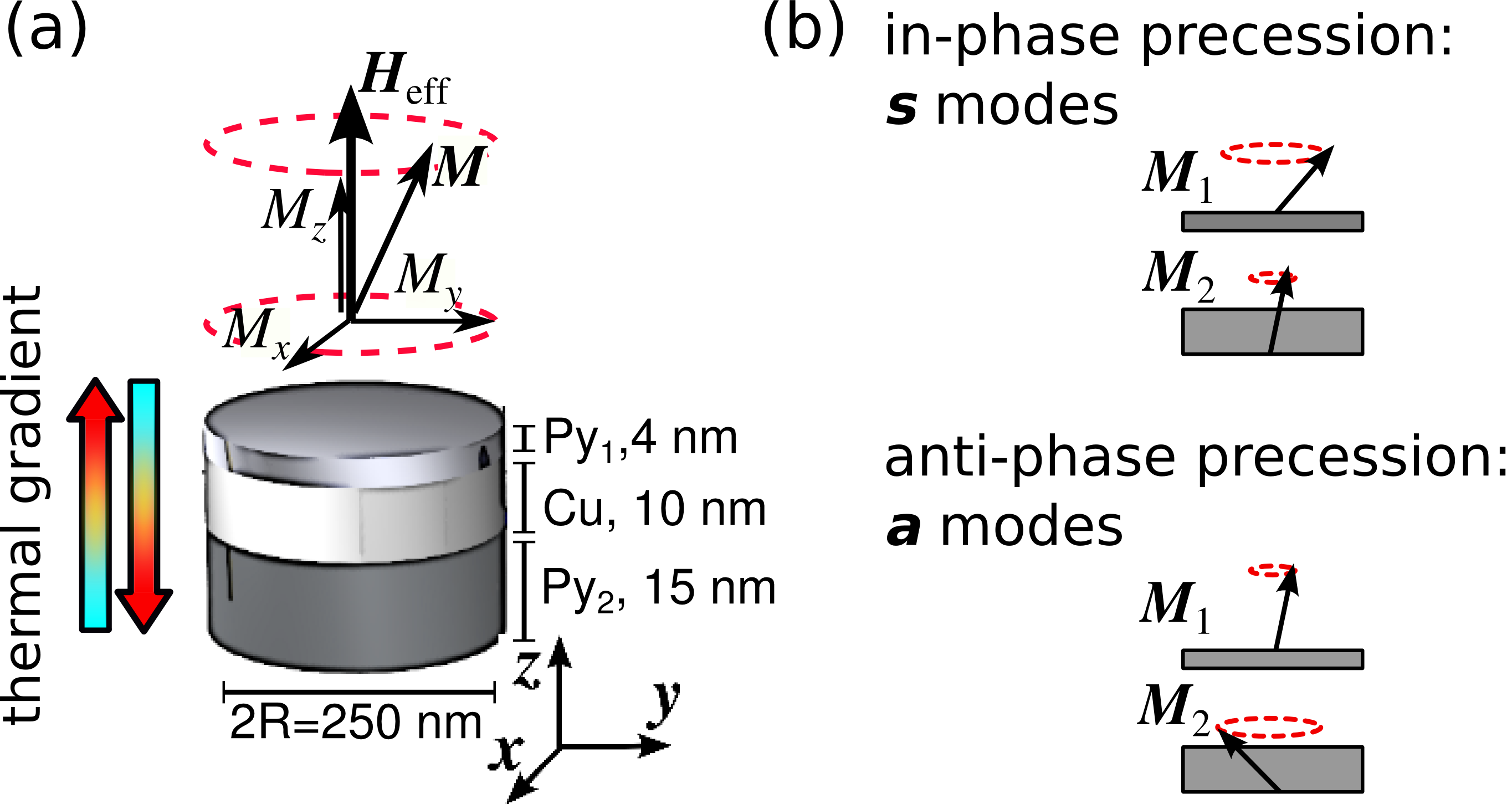}
\end{center}
\caption{(Color online).(a) Bi-layer system studied in our simulations. The magnetization is decomposed into the static component $M_z$ and the transverse components $M_x$ and $M_y)$,
which precess at the Larmor frequency in the $x-y$ plane. A uniform thermal gradient is set along the $z$ axis. (b) Symmetric ($s$) and anti-symmetric ($a$) 
precession states of the system.}
\label{fig:system}
\end{figure}

The system consists of a spin-valve nano-pillar made of two Permalloy (Py) circular disks coupled by dipolar interaction, see Fig.\ref{fig:system}(a). A uniform thermal
gradient is applied along the $z$ direction. The origin of the rectification effect, which is similar to the case of the conventional thermal diode \cite{casati04}, 
resides in the fact that the spin-wave (SW) spectra of the disks are temperature dependent and their overlap can be controlled by the gradient. 

This device has novel features with respect to both the conventional thermal diode and the spin-caloritronics devices studied so far. 
In particular the rectification of \emph{two coupled currents} is a novel phenomenon that has not 
been investigated before. Then, the SW spectra of the nano-pillar have several SW modes. Only some of those modes can overlap in the presence of a thermal gradient, leading to a 
"partial phase-locking" between the two disks. The definition of magnetization current between the two layers emerges naturally within the formulation of the problem, and
extends the notion of the usual SW-spin current \cite{kajiwara10} to discrete multi-mode systems coupled by dipolar interaction.
Finally, the present study is performed on a realistic device \cite{thesis,naletov11}, suggesting possible experimental investigations .
 
Let us start with a brief review of the magnetization dynamics inside our system, before discussing the effect of the thermal gradient.
The local dynamics of the magnetization in a ferromagnet is described, at the length scale of the exchange length, by the classical Landau-Lifshiz-Gilbert (LLG) equation of motion \cite{landau65,gilbert55} 

\be\label{eq:LLG}
\frac{\der{\bm M}}{\der{t}}= -\gamma_0 {\bm M}\times {\bm H}_{\mbox {eff}} + \frac{\alpha}{M_s}{\bm M}\times{\frac{\der{\bm M}}{\der t}}.
\ee
Here,  $\gamma_0$ is the gyromagnetic ratio, $\alpha$ is the Gilbert damping parameter and $M_s$ is the saturation magnetization of the sample. 
The first term at the rhs of Eq.(\ref{eq:LLG}) is the adiabatic torque, which describes the precession of the magnetization $\bm M$ around the effective field 
${\bm H}_{\rm{eff}}$ \cite{gurevich96}. Here, the latter contains respectively external, exchange and demagnetizing field \cite{thesis,naletov11}. The second term at the rhs of Eq.(\ref{eq:LLG}), 
describes energy dissipation at a rate proportional to $\alpha$, so that in absence of external excitations the magnetization eventually alignes with $\bm{H}_{\rm{eff}}$.

Below we briefly review the classification of the SW modes of our system, see Refs.\cite{thesis,naletov11} for a thorough discussion. 
The magnetization dynamics in a confined nano-structure is described by the continuous magnetization vector field  $\bm{M}(\bm{r},t)$.
In the case of thin layers considered here, $\bm{M}$ is uniform along the thickness, so that the LLG equation simplifies to two equations describing the
circular precession of $M_x$ and $M_y$ around $z$ \cite{slavin09}. These equations can be rewritten as one complex equation for the dimensionless 
SW amplitude $c=(M_x + i M_y)/\sqrt{2M_s(M_s+M_z)}$, which depends on the polar coordinates $(\rho,\phi)$.

The dynamics of the two disks, written in terms of $c(\rho,\phi,t)$ is described by the equations
\beA\label{eq:osc}
\dot{c}_1 & = &  i\omega_1(p_1)c_1-\Gamma_1(p_1)c_1 + ih_{12}c_{2},\\
\dot{c}_2 & = & i\omega_2(p_2)c_2-\Gamma_j(p_2)c_2 + ih_{21}c_{1},\nonumber
\eeA
which are the equations of motion of two coupled \emph{nonlinear} oscillators with resonance frequency $\omega_j(p_j)$,  $j=1,2$. The term $\Gamma_j(p_j)$ is the damping rate, 
responsible for the finite linewidth of the resonance peaks \cite{slavin09,gurevich96}. Both terms depend on the SW power $p_j=|c_j|^2$, which describes the amplitude of the oscillations in each disk \cite{slavin09}. 
The coupling strength $h_{jj^\prime}$, due the dynamical dipolar coupling between the two disks, is an effective term obtained averaging the dipolar field over the volumes of the samples, see Ref.\cite{naletov11}
for the explicit expression.

The normal SW modes of an isolated disk are given by $c_{\ell,n}(\rho,\phi,t) = J_\ell (k_{\ell,n} \rho)\exp(+ i \ell \phi)\exp(i \omega_{\ell,n} t)$
where $J_\ell$s are Bessel functions of the first kind, $\omega_{\ell,n}$ is the resonance frequency and $k_{\ell,n}$ is the norm of the in-plane SW wave
vector. In this notation, $n$ and $\ell$ represent respectively the radial and azimuthal mode index. The $\ell$ index determines the coupling of the system with an external rf source:
the $\ell=0$ modes are excited only by a uniform in-plane field, while the $\ell=1$ modes are excited only by an orthoradial field \cite{naletov11,thesis}.

In the case of two thin disks coupled via dipolar interaction considered here, the spatial profile of the SW modes is unchanged, 
while the collective magnetization dynamics separates into a bonding (or symmetric, $s$) and an anti-bonding (antisymmetric, $a$) state with different resonance frequencies.
The first corresponds to an in-phase precession of the two disks, that occurs mainly in the thin layer, while the latter corresponds to an anti-phase precession, that occurs mainly in the thick layer, 
see Fig.{\ref{fig:system}}(b) for a cartoon. 

To resume, each peak of the SW spectrum is identified as $s_{\ell,m}/a_{\ell,m}$, according to the disk which precesses the most, and to the azimuthal and radial indexes. 

\begin{figure}
\begin{center}
\includegraphics[width=6.5cm]{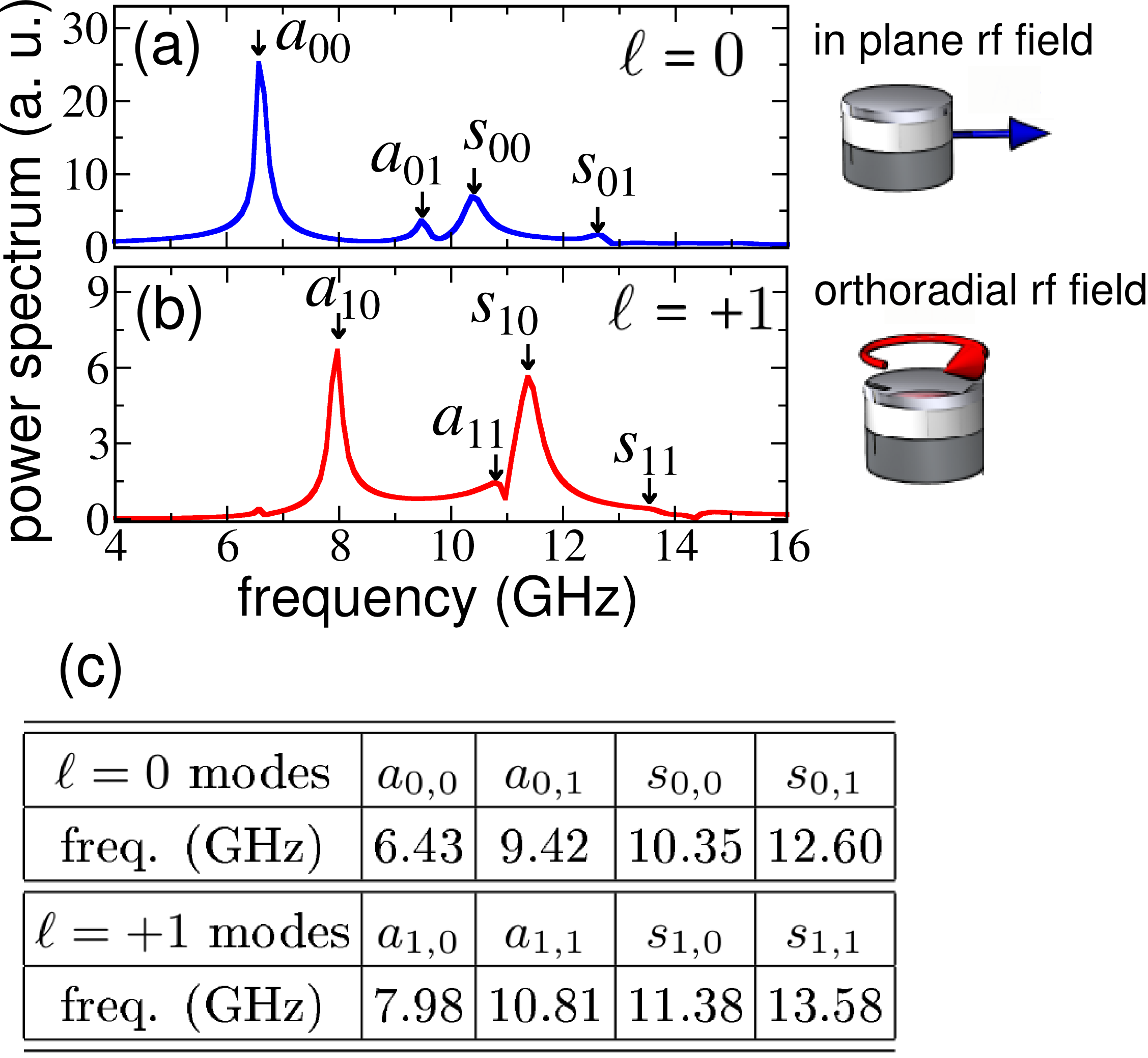}
\end{center}
\caption{(Color online). (a) and (b) SW spectrum for $\ell=0$ and $\ell=+1$ modes respectively.(c) Tabulation of the corresponding frequencies.}
\label{fig:spectrum}
\end{figure}

Let us now turn to micromagnetic simulations. The system studied, shown in Fig.\ref{fig:system}, consists of two Permalloy disks Py$_j$, $j=1,2$, separated by 10 nm. The layers have thicknesses 
$t_j$ of 4 and 15 nm correspondingly, and a radius $R=125$ nm. The exchange stiffness of Py is $A=1.3\times 10^{-11}$ J/m. The magnetic parameters of the disks, 
taken from Ref.\cite{thesis}, are $M_{s1}=7.8\times 10^5$ A/m, $M_{s2}=9.4\times 10^5$ A/m, $\alpha_1=1.6\times 10^{-2}$, $\alpha_2=0.85\times 10^{-2}$ and $\gamma_0=1.87\times 10^{11}$ rad$\times$s$^{-1}$$\times$T$^{-1}$.
The sample is saturated by an external field of 1 T in the $z$ direction.
  
The simulations were performed with the Nmag finite element micromagnetic solver \cite{fischbacher07}. The integration of the LLG equation at each mesh site
is performed by the Sundials ODE solver \cite{sundials}, which is based on variable steps multistep methods. The tetrahedral mesh, automatically generated by the Netgen package \cite{netgen}, 
has a maximum intersite distance of 6 nm, of the order of the Py exchange length. 

Thermal fluctuations are introduced by adding to the effective field ${\bm H}_{\rm{eff}}^k$ at site $k$ of the mesh, a stochastic field ${\bm H}_{\rm{th}}^k$. The latter is assumed to be a Gaussian random process 
with zero mean and amplitude $\average{{\bm H}_{\rm{th},i}^k{\bm H}_{\rm{th},j}^l}=2D_k\delta_{ij}\delta_{kl}\delta(t-t')$. Here $i,j=x,y,z$ stand for the cartesian 
components of the field, while $k,l$ refers to the sites on the mesh. The fluctuation amplitude is  
$D_k=(2\alpha k_B T_k)/(M_s\gamma_0 \mu_0 V_k)$, where $k_B$ is the Boltzmann constant, $\mu_0$ is the vacuum magnetic permeability, $T_k$ is the temperature at site $k$ and
$V_k$ is the volume containing the magnetic moment at site $k$ \cite{martinez07}. In Permalloy, the parameter $\alpha$ does not depend on the temperature \cite{lubitz01}.

The quantity of interest in our simulations is the space-averaged magnetization $\average{\bm{M}_j(t)}=\frac{1}{V_j}\int_{V_j}{{\bm{M_j}}(\bm{r},t){\rm{d}}^3{\bm{r}}}$ of the disk $j=1,2$, 
which is used to compute the SW amplitude $c_j$. The power spectrum is computed from the collective SW amplitude averaged over the sample thicknesses \cite{naletov11}:  $c=(c_1t_1+c_2t_2)/(t_1+t_2)$.

The $\ell=0$ modes (displayed in blue tones) are excited starting from an initial condition  where the magnetization uniformly tilted of $8^\circ$ in the $x$ direction with respect to the precession axis $z$. 
The $\ell=+1$ modes (displayed in red tones) are excited applying to the magnetization aligned with the $z$ axis the perturbation orthoradial vector field $\bm{\theta}(\rho,z)=\epsilon \hat{\bm{z}}\times\hat{\bm{\rho}}$, 
where $\epsilon=0.5$ and $\hat{\bm{\rho}}$ is the unit vector in the radial direction.
Starting from these conditions, we have computed the time evolution of the system for 50 ns, with an integration time step of 1 ps. 

Figures {\ref{fig:spectrum}} (a) and (b) show the power spectrum of the system at zero temperature.
The frequencies of the peaks, tabulated in Fig.\ref{fig:spectrum}(c), agree with Refs.\cite{thesis,naletov11}. 
The relative height of the peaks depends on the initial conditions, which in our case favour the low frequency modes ($a_{00},a_{10},s_{00},s_{10}$).
We focus on the analysis of those modes, that dominate the spectrum. 

\begin{figure}
\begin{center}
\includegraphics[width=8.5cm]{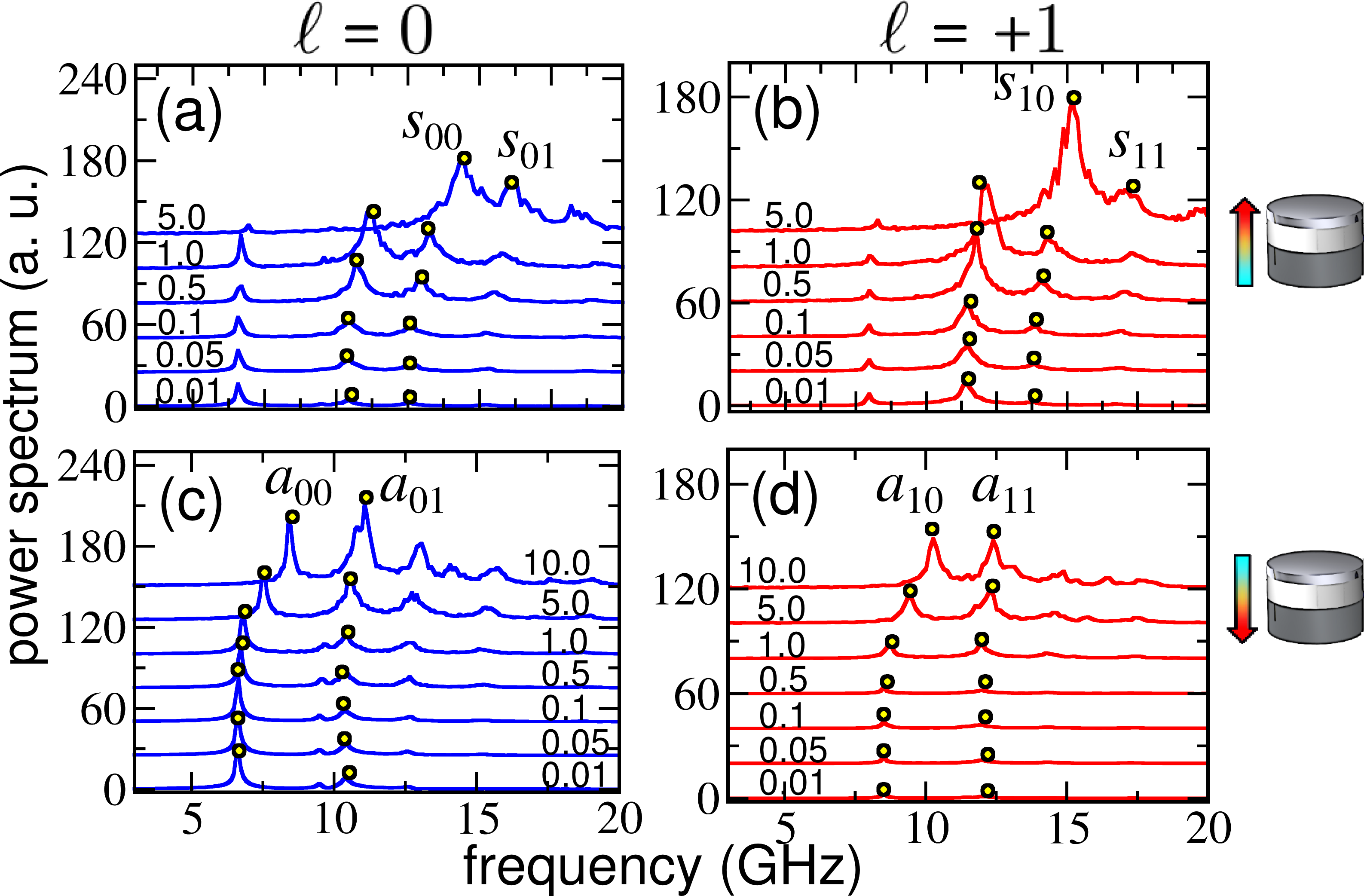}
\end{center}
\caption{(Color online). SW spectrum for different $\ell$ modes in the presence of thermal gradient (expressed in K/nm). 
(a) and (b): positive gradient, (c) and (d) negative gradient. Positive (resp.) negative gradients excite only the $s$ (resp. $a$) modes.}
\label{fig:big_spectrum}
\end{figure}

Let us now turn to the description of the system in the presence of the temperature gradient.
We consider the gradient positive when the temperature increases along $z$ (thin layer hotter than the thick one) and negative in the opposite case. 
The low temperature side of the disks is always kept at 0 K.
The computatons at finite temperature were averaged over 24 samples with different realization of the disorder. 

Figures \ref{fig:big_spectrum} (a) and (b) show the effect of a positive gradient on the modes with $\ell=0$ and $\ell=+1$ correspondingly.
When the gradient is positive, only the symmetric modes $s_{\ell,n}$ are excited. They grow in height starting from $+10^{-2}$ K/m, and eventually dominate the spectrum. 
Between 0.5 and 1 K/nm, those modes shift towards higher frequencies. This effect is typical of nonlinear oscillators, where the frequency depends on the oscillation amplitude \cite{slavin09}. 
Notice that the anti-symmetric modes $a_{\ell,n}$ do not modify their frequency, while their amplitude remains constant until 1 K/nm and then decreases. 

Figures \ref{fig:big_spectrum} (c) and (d) show the effect obtained reversing the gradient, where only the anti-symmetric modes grow and shift towards higher frequency.
Notice that in Fig.\ref{fig:big_spectrum}(c) the modes $a_{01}$ and $s_{01}$ are very close in frequency and merge in a single mode at high gradient. 

\begin{figure}
\begin{center}
\includegraphics[width=5.5cm]{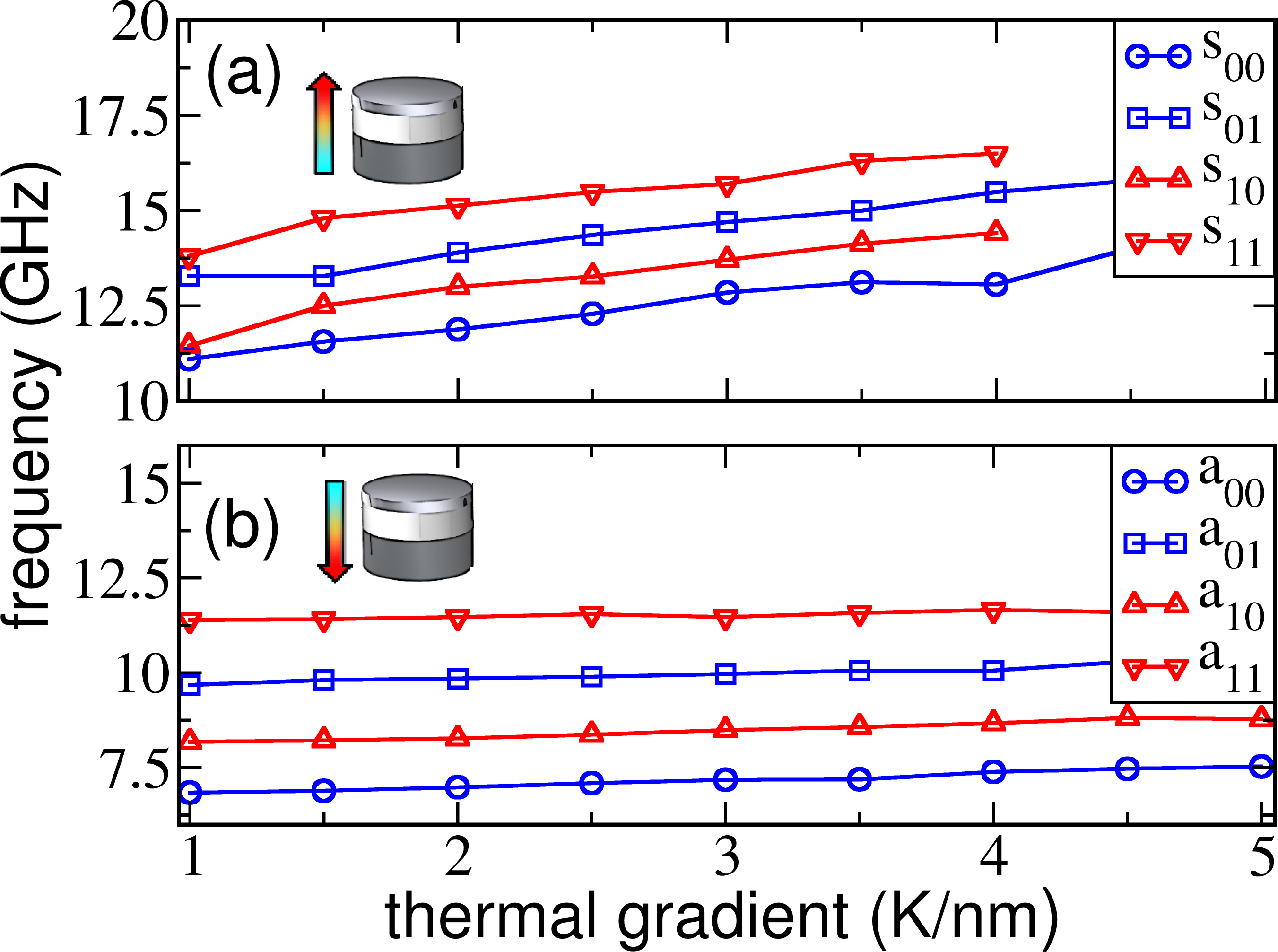}
\end{center}
\caption{(Color online). Frequency of the SW modes as a function of a positive (a) and a negative (b) gradient. The lines are guide to the eye.}
\label{fig:modes}
\end{figure}

The frequencies of the excited SW modes increases roughly linearly as a function of the gradient, see Figs.\ref{fig:modes} (a) and (b). 

We discuss now the main result of the paper, that is, the capability of the system to operate as a diode. We start by introducing the Hamiltonian of the problem

\be\label{eq:hamiltonian}
\mathcal{H}=\omega_1(p_1)p_1+\omega_2(p_2)p_2+h(c_1c_2^*+c_1^*c_2),
\ee
where for simplicity we have taken a \emph{symmetric} coupling $h_{12}=h_{21}=h$. Equation (\ref{eq:hamiltonian}) leads to the conservative part of Eqs.(\ref{eq:osc}) through $\dot{c}_j=i\delta\mathcal{H}/\delta c_j^*$ 
\cite{naletov11,slavin09,iubini13}. Notice that Eq.(\ref{eq:hamiltonian}) is the Hamiltonian of a nonlinear \emph{Schroedinger dimer} (SD), the simplest possible realization of 
the discrete non-linear Schroedinger (DNLS) chain. The DNLS, which has many applications in other branches of physics \cite{kevrekidis01,eilbeck03,flach08},
it is also used to model the small amplitude dynamics of a spin chain and, in its continuum version, the SW propagation in ferromagnets \cite{rumpf03}. 

It is known that the SD (and in general the DNLS) have two conserved quantities, to which correspond two conserved currents \cite{rumpf03,iubini12,iubini13}: the total "number of particle" 
(in our case, the SW power $p_1+p_2$) and the total energy $\mathcal{H}$. Multiplying Eqs.(\ref{eq:osc}) by their complex conjugate and summing them, as in Ref.\cite{slavin09}, 
gives the conservation equation for the SW power 

\be\label{eq:swpower}
\dot{p}_1 = -2\Gamma_1(p_1)p_1+j_{\bm{M}},
\ee
and a similar equation for $p_2$. This leads to the definition of the magnetization current $j_{\bm{M}}=2h\mbox{Im}(c_1c_2^*)$ between the two oscillators 

The energy current $j_E$ is implicitly defined by the conservation equation for the \emph{local} energy \cite{lepri03,rumpf03,iubini12,iubini13}: 

\be\label{eq:ecurrent}
\dot{\mathcal{H}}_2=j_E,
\ee
where $\mathcal{H}_{\rm{2}}=p_2+h(c_1c_2^*+c_1^*c_2)$ is the energy of oscillator 2.
An explicit calculation using Eqs.(\ref{eq:osc}) gives $j_E=2h{\rm{Re}}(\dot{c}_1c_2^*)$.
Notice that those currents are obtained in the same way as the currents for the DNLS chain \cite{lepri03,iubini12,iubini13}, and they represent the special case where the DNLS has only two elements.
In particular, Eq.\ref{eq:swpower}) constitutes the continuity equations for the $z$ component of the magnetization in each disk, and in a continuum ferromagnet leads to the usual definition of SW current \cite{rumpf03}.

Notice that, if $h_{12}\neq h_{21}$, the Hamiltonian is not real and the total SW power is not conserved. This issue is resolved simply by
rescaling the SW amplitudes in Eqs.(\ref{eq:osc}) as $c_1\rightarrow\sqrt{h_{12}}c_1$ and $c_2\rightarrow\sqrt{h_{21}}c_2$ and leads to the change $h\rightarrow\sqrt{h_{12}h_{21}}$ in the currents. 

%
\begin{figure}
\begin{center}
\includegraphics[width=7.0cm]{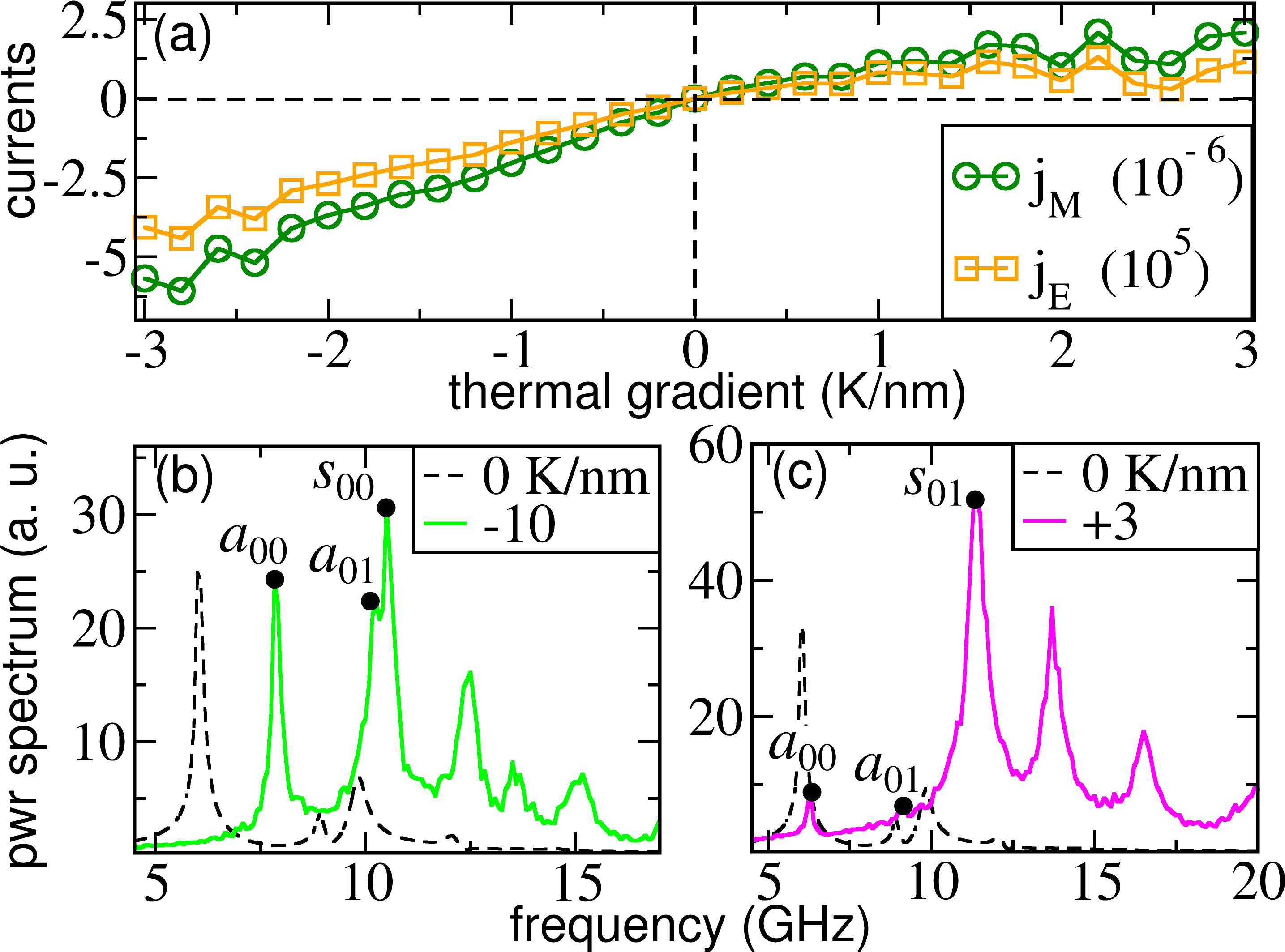}
\end{center}
\caption{(Color online). (a) Rectification effect for magnetization and energy currents. The panels (b) and (c) show the powers spectra computed for different gradients and illustrate the rectification effect.
(b) For negative gradients, the modes $a_{01}$ and $s_{01}$ overlap, giving a partial phase-locking and a conductive state. (c) For positive gradients, there is no overlapping and the conduction is reduced.}
\label{fig:currents}
\end{figure}

Let us now consider the numerical simulations. The currents where computed from the time evolution for the $\ell=0$ modes, in the gradient interval between $\pm 3$ K/nm, near the linear regime.  
The currents displayed in our figures are the correlation functions $j_M={\rm{Im}}\average{c_1c_2^*}$ and $j_E=\rm{Re}\average{\dot{c}_2c_1^*}$, where $\average{\cdot}$ 
stands for both the ensemble and the time average. The latter is performed in the interval between 20 and 50 ns. Fig.\ref{fig:currents}(a) shows the two currents as a function of the thermal gradient. One can see that the system displays a rectification effect, with the conducting  state at negative gradients.

This effect can be explained in a way similar to the conventional thermal diode \cite{casati04}. 
For an oscillator with only one SW mode, the SW amplitude can be written in the phase-amplitude representation as $c_{j}=\sqrt{p_j}e^{i\phi_j}$, for $j=1,2$.
The currents then read $j_{\bm{M}}=2h\sqrt{p_1p_2}{\sin}(\Delta\phi)$ and $j_E=2h\sqrt{p_1p_2}\omega_1\mbox{sin}(\Delta\phi)$, with $\Delta\phi=\phi_1-\phi_2$ and $\dot{\phi}_j=\omega_j$. 
At zero gradient, the two oscillators precess with different frequencies, so that the two phases $\phi_1$ and $\phi_2$ increase with different 
rates. Thus the currents oscillate in time with zero average value. In this free-running phase configuration the system is in the insulating state. The crucial point is that, since the system is nonlinear, the frequencies 
are temperature dependent. Thus, in the presence of a gradient the spectra may overlap, so that $\Delta\phi$ becomes constant. In this phase-locked configuration the currents are constant and the system behaves as a conductor.

The phase locking can be clearly seen in Fig.\ref{fig:currents}(b). In the presence of a negative gradient, all the $a$ modes shift towards higher frequencies
and approach the $s$ modes. In particular $a_{01}$ and $s_{00}$ \emph{merge into a single mode}. On the contrary, with a positive gradient the $s$ modes are the ones that shift towards higher frequencies,
so that the frequency gap between the two oscillators increases.

In the multi-mode system considered here, the magnetization current reads $j_{\bm{M}}~=~2h{\mbox{Im}}(\sum_{\ell,m,\ell^\prime,m^\prime}\average{a_{\ell,m} s_{\ell^\prime,m^\prime}^*})$, 
and a similar expression holds for the energy current. In the conducting state, $j_{\bm{M}}$ consists of a sum of oscillating and constant signals, the latter corresponding to the modes that
overlap. Thus, its power spectrum is a sequence of lorentzian peaks, wich include a zero frequency mode that accounts for the constant components of the current. 

\begin{figure}
\begin{center}
\includegraphics[width=7.0cm]{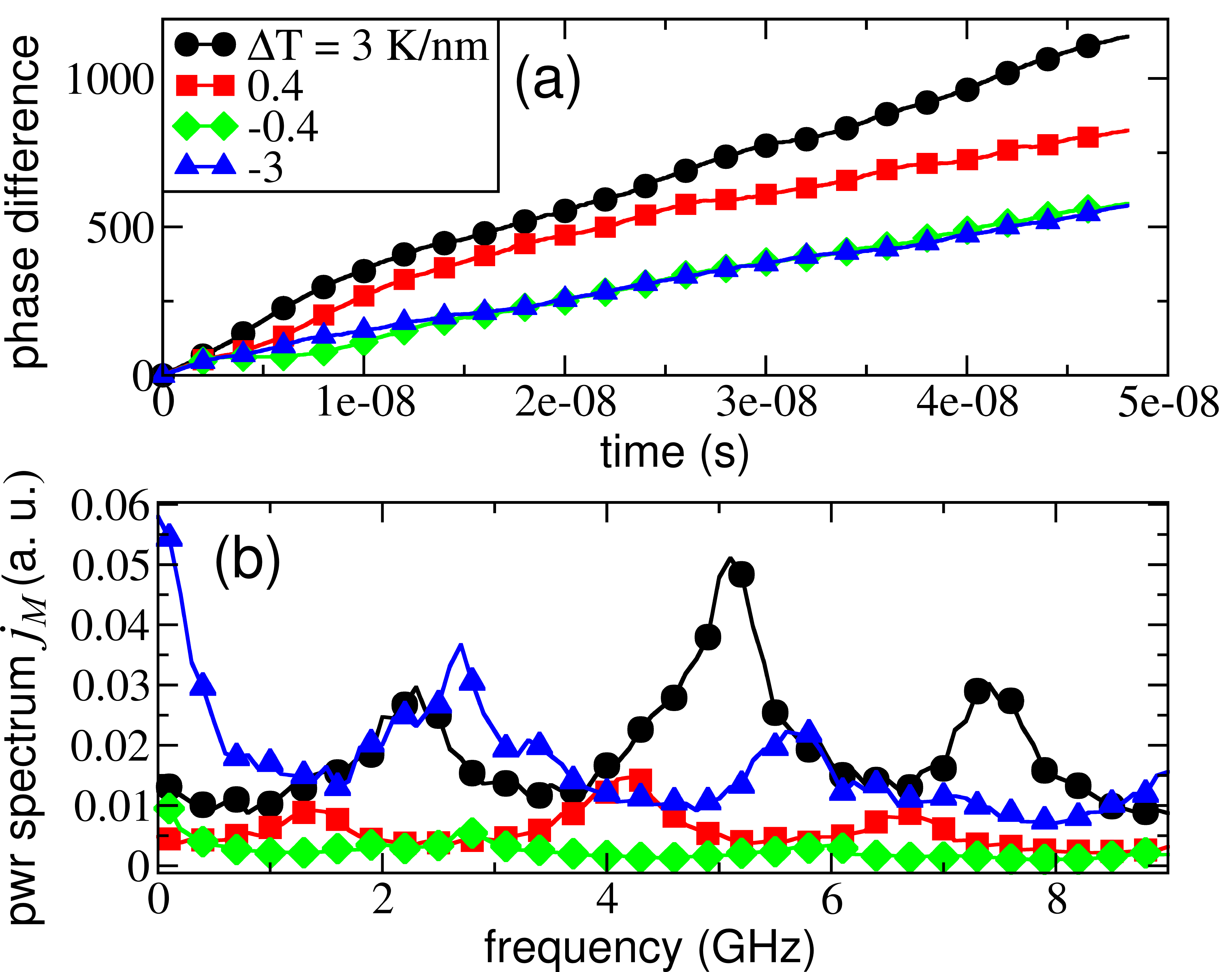}
\end{center}
\caption{(Color online). (a) Phase difference $\Delta\phi$ between the oscillators vs time, computed for different thermal gradients. $\Delta\phi$ increases faster in the insulating region than in the conducting one,
where the frequencies of the two oscillators become closer. (b) Corresponding power spectrum of the magnetization currents. In the conductive state, the spectrum is dominated by the zero frequency mode,
which is barely visible in the insulating state. In the insulating state, the spectra are reduced by a factor 3 for better visibility.}
\label{fig:phases}
\end{figure}

This physical picture is supported by Fig.\ref{fig:phases}. Panel (a) shows $\Delta\phi$ as a function of time for different thermal gradient. One can see that $\Delta\phi$ increases 
much faster in the insulating region than in the conductive one, indicating that in the latter case the frequencies of the two oscillators become closer. Notice that, if the overlap of the modes were complete,
$\Delta\phi$ would simply fluctuate around zero because of thermal noise. Panel (b) shows the power spectrum of $j_M$ as a function of frequency, computed for different gradients. One can clearly see 
the difference between the conducting and the insulating state: In the first case, the spectrum is dominated by the zero frequency, while in the latter the zero mode is not visible and the spectrum is
dominated by the finite frequency modes. 

To conclude, we have performed a numerical study that describes a novel phenomenon: the rectification effect of both energy and magnetization currents in a realistic spin-valve system with several SW modes.
The description is based on the natural extension of the concept of SW-spin current to a discrete system coupled by dipolar interaction.  
The definition of magnetization currents as correlation functions between SW amplitudes suggests a new way to measure spin currents in spin-Seebeck devices.
We gratefully acknowledge financial support from Carl Tryggers Stiftelse (CTS), 
Goran Gustafssons stiftelse, 
Vetenskapradet (VR), the Royal Swedish Academy of Sciences (KVA), the Knut and Alice Wal
lenberg foundation (KAW), the European Commission (NexTec project), and the Swedish Foundation for Strategic Research (NanoTEG project). 
The computer simulations were performed on resources provided by the Swedish National Infrastructure for Computing (SNIC) at National Supercomputer Centre (NSC). 
We thank S. Lepri, S. Iubini and M. Molinari for useful discussions.

\end{document}